\documentclass[12pt]{article}
\usepackage[a4paper,margin=1in]{geometry}
\usepackage{graphicx}
\usepackage{caption}
\usepackage{titlesec}
\usepackage[hyphens]{url}
\usepackage{hyperref}
\hypersetup{breaklinks=true}
\usepackage{textcomp}
\usepackage{parskip}
\titleformat{\section}{\large\bfseries}{\thesection}{1em}{}
\titleformat{\subsection}{\normalsize\bfseries}{\thesubsection}{1em}{}

\title{Astronomical Refutation of the New Chronology by Fomenko and Nosovsky:\\
The 1151-Year Planetary Cycle and Dating of the Almagest via Speed/Error Correlation}
\author{Carlos Baiget Orts\\
\small Computer engineer, independent researcher in historical astronomy\\
\small \texttt{asinfreedom@gmail.com}}
\date{}

\begin{document}
\pagenumbering{gobble}  % Oculta la numeración en la portada
\maketitle

\begin{abstract}
Two astronomical methods developed through computational simulation are presented, enabling historical dating from a technical and reproducible perspective. The first identifies a 1151-year astronomical cycle based on the recurrence of visible planetary configurations (from Mercury to Saturn), along with the Sun and Moon—all considered planets within the framework of the ancient geocentric solar system. This cycle was detected by an algorithm that analyzes ephemerides from a geocentric perspective using the Skyfield astronomy library \cite{skyfield}.

The second method, named \textbf{SESCC} (Speed-Error Signals Cross Correlation), is a statistical procedure for dating ancient star catalogs, with a particular focus on Ptolemy’s Almagest. It relies on the correlation between the positional error of stars and their proper motion in ecliptic latitude. This correlation reaches a minimum when the catalog is compared to its actual compilation date, because at that epoch, the apparent displacements caused by proper motion have not yet accumulated. Thus, fast-moving stars do not systematically deviate more than slower ones, eliminating any statistical correlation between error and velocity. This behavior allows for an objective and reproducible estimate of the epoch.

Both methods challenge key pillars of the New Chronology by Fomenko and Nosovsky. The source code of the algorithms is publicly available, facilitating both reproducibility and independent evaluation.

\end{abstract}
\newpage
\pagenumbering{arabic}  % Restaura numeración norma

\section{Introduction}
The New Chronology, formulated by Anatoly Fomenko and Gleb Nosovsky \cite{fomenko2003}, proposes a radical revision of ancient and medieval history. It argues that much of the conventional chronology is the result of falsifications or systematic errors. This theory relies, in part, on reinterpretations of ancient astronomical data, such as horoscopes and star catalogs.

This work presents two algorithms developed at the end of 2023 and validated during 2024, designed to precisely evaluate historical astronomical datings. 

Both have been applied to key sources such as Ptolemy’s \textit{Almagest} and yield results that directly conflict with two fundamental corollaries of the New Chronology: the proposed dating of the Anno Domini to the year 1152—which is precisely 1151 years after the traditional year 1—and the claim that prehistory ended around the 11th century \cite{fomenko2012}. Moreover, the New Chronology places the Crucifixion in the year 1185, exactly 1151 years after the conventional date of 34 CE. These numerical correspondences with the 1151-year planetary cycle raise the possibility that such shifts may stem from astronomical ambiguities rather than historical reality.

\section{The Ancient Solar System and the 1151-Year Cycle}

\begin{figure}[h]
\centering
\includegraphics[width=0.85\textwidth]{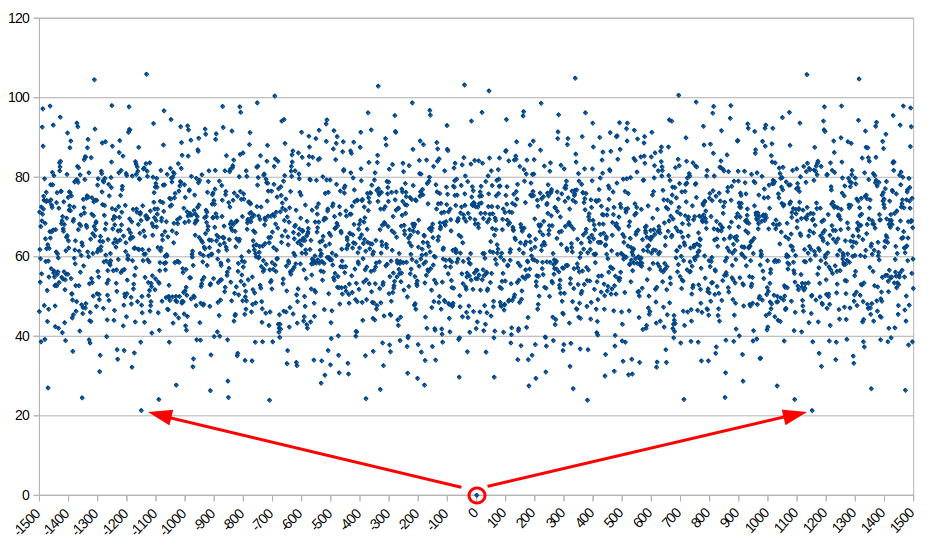}
\caption{Scatter plot generated by the 1151-year cycle algorithm. The X-axis represents years from the reference date, and the Y-axis shows the mean angular deviation. Clear minima appear at ±1151 years, indicating a stable cyclical recurrence of planetary configurations.}
\end{figure}

A visual animation illustrating this phenomenon is available at: \href{https://www.youtube.com/watch?v=W4z_anTXi9U}{1151-Year Cycle Visualization}.

In antiquity, the solar system was conceived from a geocentric perspective, comprising the five visible planets (Mercury, Venus, Mars, Jupiter, and Saturn), along with the Sun and the Moon. All these bodies were considered wandering stars or “planets.” The search for repeating patterns in their configurations led to the development of an algorithm that analyzes time series of planetary positions using the Skyfield library.

This cycle was not only clearly identified, but also emerged as the best among all candidate cycles, based on the combined criterion of minimum average deviation and low dispersion, where 'low dispersion' refers to the standard deviation of the angular displacements across the planets, indicating uniform displacement rather than irregular or inconsistent offsets. 

No other time interval exhibited a comparable level of sustained similarity, reinforcing its uniqueness.

The algorithm calculates the angular arrangement of these celestial bodies relative to a reference date and compares each day in a sequence (or 'series') of dates before and after that reference. Each candidate series produces a mean angular deviation, and the algorithm searches for sustained periods of minimal deviation. A stable coincidence in the general configuration was identified with a periodicity of exactly 420403 days, equivalent to 1151 years.

\section{Methodology of the Cyclical Analysis}

The analysis of the 1151-year cycle is based on the systematic comparison of the angular arrangement of the seven classical planets (from Saturn to the Moon), from a geocentric perspective. Positions are obtained in ecliptic coordinates under the J2000 reference system, covering the period from the year -1500 to 1950, with daily resolution. The data used consist of the ecliptic longitudes of the seven celestial bodies considered “planets” in antiquity: the Sun, Moon, Saturn, Jupiter, Mars, Venus, and Mercury. These data are stored in a structured file precomputed using the Skyfield astronomy library \cite{skyfield}.

Each comparison series is defined from a reference date and has a configurable duration: from one year to several centuries. For each candidate date, either before or after the reference date, the mean angular displacement between each planet's position and its counterpart is calculated, along with the standard deviation of these differences.

Angular distances are measured in absolute value and adjusted using circular symmetry: if a difference exceeds 180º, it is replaced by its supplementary angle (360º minus the difference). This avoids discontinuities inherent to angular systems. The resulting values for each candidate cycle are aggregated by summing the mean deviation and its standard deviation, which is valid since both quantities are expressed in degrees and represent comparable angular magnitudes. The best cycle is the one that minimizes this sum: a low mean deviation indicates similarity, and a low standard deviation implies uniformity across planets.

The program executes efficiently thanks to the use of a compressed file containing precomputed planetary positions. For each reference date, the program analyzes all possible cycles within the defined search interval and generates a scatter plot. In this plot, the x-axis represents years, and the y-axis the mean deviation associated with each year relative to the reference date. Results show that dates located at ±1151 years (equivalent to ±420403 days) from the base date display the lowest values on the graph, with an average deviation of about 21 degrees and consistently low standard deviation.

This pattern holds for any reference date and any series length, reinforcing the stability and uniqueness of the cycle. The periodicity is not influenced by the observer’s location on Earth, as the model is entirely geocentric. 

Although no dynamic explanation has been proposed for this phenomenon, its existence is empirically demonstrated through a recognized astronomical calculation model. Interestingly, multi-body orbital resonances and synchronizations have been observed in exoplanetary systems, such as the six-planet resonance chain discovered by the TESS mission \cite{tess2021resonance}. These findings show that highly coordinated planetary dynamics are not only possible but may occur naturally under certain conditions, suggesting that the temporal symmetry observed in the 1151-year cycle could warrant further investigation from a dynamical perspective.

In addition to the recurrence of a similar planetary configuration with a consistent angular displacement, the apparent retrograde motions of the planets also exhibit synchronization across the 1151-year intervals. These complex, periodic reversals—as seen in geocentric perspective—further reinforce the robustness and internal consistency of the cycle.

The most significant implication of this finding is that any attempt to date ancient horoscopes without accounting for this cycle may lead to systematic errors exceeding a millennium.

\section{Dating Star Catalogs Using SESCC (Speed-Error Signals Cross Correlation)}

\begin{figure}[h]
\centering
\includegraphics[width=0.85\textwidth]{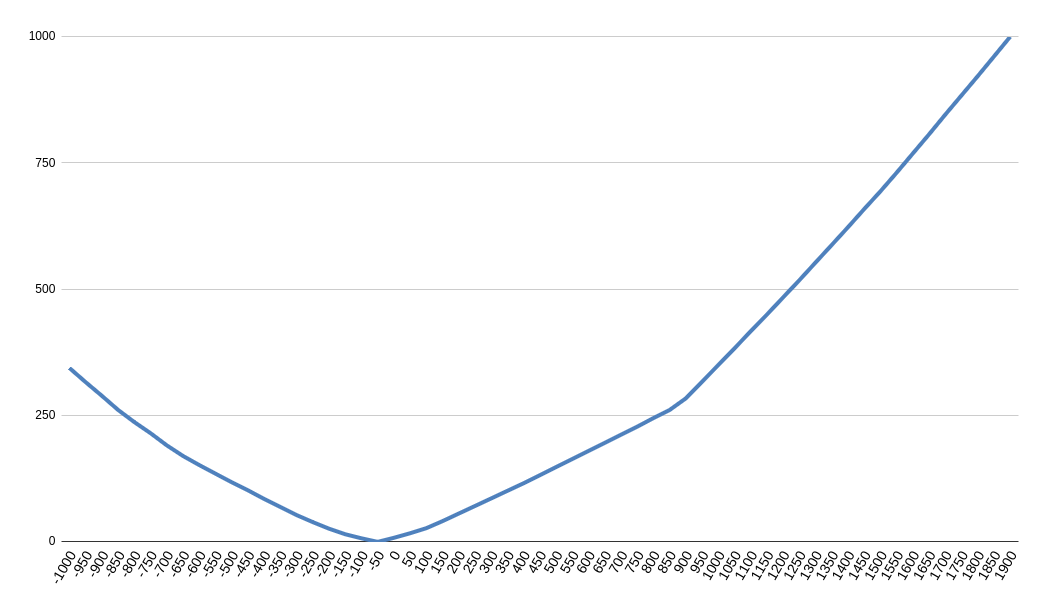}
\caption{Correlation curve generated by the SESCC method applied to the \textit{Almagest}. The minimum around year -50 indicates the most probable observation date.}
\end{figure}

This second algorithm is based on a simple yet powerful premise: positional errors in the ecliptic latitudes of an ancient star catalog should not correlate with the magnitude of the stars’ proper motion. When such correlation appears, it indicates that the comparison is being made with an incorrect date. The correct date is the one where this correlation reaches a minimum, since the apparent displacements caused by proper motion have not yet accumulated significantly. As a result, fast-moving stars do not systematically deviate more than slower ones, eliminating the statistical correlation.

In this context, both the list of positional errors and the list of latitudinal velocities are interpreted as discrete signals—numerical sequences indexed by the stars in the catalog. This framing justifies the term “signals” in the method’s name. The dot product between these vectors quantifies their zero-lag cross-correlation and reveals the degree to which the error pattern aligns with the velocity distribution at a given epoch.

To implement this principle, current stellar positions derived from the Skyfield library and the Hipparcos database \cite{esa1997} are used, along with high-precision ephemerides from NASA \cite{de441}. The latitudinal velocity of each star is computed as the positional difference between two epochs. This allows generating a velocity vector for each star that can be directly compared against its positional error.

The correlation value is normalized on a scale from 0 to 1000 to enhance graphical interpretation; this normalization preserves relative correlation strengths while making minima easier to detect visually. A graph is generated showing this value across a range of candidate dates, and the minimum of the curve marks the most probable epoch of the catalog.

The method is robust against noise, identification errors, or anomalous data points. It operates directly on the raw catalog without the need for filtering. Its neutrality ensures that results are not biased by manual selection or prior assumptions.

The validity of the method has been confirmed through the generation of synthetic catalogs and their successful dating, as well as by its application to Tycho Brahe’s catalog—where the estimated date deviates by less than 50 years from the known observational period. When applied to the full \textit{Almagest} catalog (over one thousand stars), the method consistently yields a minimum around the 1\textsuperscript{st} century\,BCE, in agreement with traditional chronology.

Although the SESCC method is primarily based on ecliptic latitudes to avoid the influence of axial precession, complementary experiments have been conducted using longitudes adjusted relative to a fixed reference star, Delta Geminorum. This approach circumvents the shifting equinoctial reference and allows for meaningful comparisons across epochs. The dating results obtained through this longitude-based variant were consistent with those derived from latitudes, reinforcing the method’s reliability.

Additionally, an experimental technique based on comparing the angular distances between stars—as a sort of “chronological signature” of the catalog—yielded similar results. Although this line of investigation requires further refinement, especially in modeling longitudinal velocities, the initial agreement across methods suggests that the \textit{Almagest} can also be reliably dated using multiple independent astronomical metrics. Detailed explanations and code implementations are available in the README of the SESCC repository.

\section{Application to the Almagest}

The SESCC method has been applied to Ptolemy’s \textit{Almagest}, the classical star catalog containing more than one thousand entries in ecliptic coordinates. Using identifications based on Toomer's reconstruction \cite{toomer1998} and positions calculated with the Skyfield library \cite{skyfield}, the algorithm was able to determine the epoch that best fits the internal structure of the catalog.

With a time resolution of 100 years and no data filtering, the method consistently identifies a minimum around the 1\textsuperscript{st} century\,BCE. This result is stable even when using randomly selected subsets of the catalog or excluding ambiguous entries, reinforcing the robustness of the procedure.

This dating aligns with the traditionally accepted chronology and directly contradicts the hypothesis put forward by the New Chronology, which places the compilation of the \textit{Almagest} sometime between the 7th and 13th centuries CE. The results confirm that the structure of the catalog reflects accurate positional knowledge consistent with the astronomical capabilities of the classical era.

Additionally, the same procedure has been successfully applied to Tycho Brahe’s catalog, yielding a result within 50 years of his historically documented observational period, further validating the general applicability and accuracy of the SESCC method.

\section{Conclusions}

Both methods offer a reproducible, data-driven challenge to historical reconstructions based solely on textual interpretation.

Further investigation has revealed that the software tool "HOROS", developed and used by proponents of the New Chronology, is parametrized in a way that systematically favors late datings. Specifically, the zodiac boundaries encoded in the program are shifted toward higher ecliptic longitudes—initially placing Aries at 26º, and later revised in 2007 to 31º. Combined with an internal tolerance of ±5°, this results in potential misalignments of up to 10°, which is sufficient to trigger false matches across 1151-year cycles.

This effect was empirically confirmed in the construction of a unique case: a horoscope whose planetary configuration matched both 1\,BCE (traditionally considered the birthdate of Christ) and 1152 CE (the supposed birthdate of Andronicus, according to the New Chronology). When input into HOROS, both dates were returned as valid. 
This ‘double horoscope’ was considered notable by New Chronology researchers during early private correspondence. A second example is the horoscope preserved in the Leiden Aratea codex, traditionally dated to around 816 CE, but misdated by HOROS to 1433 CE. After reproducing HOROS’s results with custom software and adjusting the zodiac limits by subtracting 10°, the correct historical date was recovered.

These findings suggest that the dating errors in the New Chronology may stem not from astronomical evidence, but from model parameters tuned to align with pre-established hypotheses. Full details and source code are available in the README files of the corresponding repositories.

Additional materials are available:
\begin{itemize}
  \item \url{https://www.youtube.com/watch?v=W4z_anTXi9U}
  \item Spreadsheet: \url{https://docs.google.com/spreadsheets/d/1HH0t8uKxqkWCGXAGHsUzvAg4OmDGfDNzwau8je0OeSI}
  \item Jupyter Notebook: \url{https://mybinder.org/v2/gh/carbaior/1151cycle/HEAD?filepath=1151_years_cycle.ipynb}
\end{itemize}

Repositories:
\begin{itemize}
  \item \url{https://github.com/carbaior/1151cycle}
  \item \url{https://github.com/carbaior/sescc}
\end{itemize}
Each repository includes open-source code, a README file with documentation, and reproducible examples to independently run and verify the results.

YouTube: \textbf{@nueva\_ilusologia} 

Email: \texttt{asinfreedom@gmail.com}

\vspace{2em}
\noindent \textcopyright~
Carlos Baiget Orts. This work is licensed under a \href{https://creativecommons.org/licenses/by/4.0/}{Creative Commons Attribution 4.0 International License (CC BY 4.0)}.


\begin{thebibliography}{99}

\bibitem{fomenko2003}
Fomenko, A. T., Kalashnikov, V. V., \& Nosovsky, G. V. (2003). \textit{History: Fiction or Science? Chronology 3}. Delamere Resources LLC.

\bibitem{fomenko2012}
Fomenko, A. T., \& Nosovsky, G. V. (2012). \textit{How It Was in Reality: Reconstruction}. Available at: \url{https://chronologia.org/en/how_it_was/index.html}

\bibitem{toomer1998}
Toomer, G. J. (1998). \textit{Ptolemy’s Almagest}. Princeton University Press.

\bibitem{esa1997}
ESA. (1997). \textit{The Hipparcos and Tycho Catalogues}. ESA SP-1200.

\bibitem{skyfield}
Rhodes, B. (n.d.). Skyfield Astronomy Library. \url{https://rhodesmill.org/skyfield/}

\bibitem{de441}
JPL (2020). DE441 Planetary Ephemerides Kernel. \url{https://naif.jpl.nasa.gov/pub/naif/generic_kernels/spk/planets/}

\bibitem{tess2021resonance}
NASA (2021). Watch the Synchronized Dance of a 6-Planet System. \textit{NASA Science – TESS Mission News}. \url{https://science.nasa.gov/missions/tess/discovery-alert-watch-the-synchronized-dance-of-a-6-planet-system/}

\end{thebibliography}
\end{document}